\documentclass[pra,twocolumn]{revtex4}
\usepackage{graphicx}
\begin{document}

\title{Dynamics of a Raman coupled model: entanglement and quantum computation}
\author{J. Larson}
\affiliation{Physics
  Department, Royal Institute of Technology (KTH)\\ Albanova,
  Roslagstullsbacken 21, SE-10691 Stockholm, Sweden} 
\author{B. M. Garraway}
\affiliation{Department of
  Physics and Astronomy\\ University of Sussex, Falmer, Brighton, BN1 9QH,
  UK}


\begin{abstract}  
  The evolution of a Raman coupled three-level $\Lambda$ atom with two
  quantized cavity modes is studied in the large detuning case; i.e.\ when the upper
  atomic level can be adiabatically eliminated. Particularly the situation
  when the two modes are prepared in initial coherent or squeezed states,
  with a large average number of photons, is investigated. It is found that
  the atom, after specific interaction times, disentangles from the two
  modes, leaving them, in certain cases, in entangled Schr\"odinger cat
  states. These disentanglement times can be controlled by adjusting the
  ratio between average numbers of photons in the two modes. It is also
  shown how this effective model may be used for implementing quantum
  information processing. Especially it is demonstrated how to generate
  various entangled states, such as EPR- and GHZ-states, and quantum logic
  operations, such as the control-not and the phase-gate.
\end{abstract}
\maketitle

\section{Introduction}
\label{sec:intro}
Entanglement is one of the most striking features of quantum mechanics. It
is fundamental for non-locality of quantum mechanics, quantum computing and
information processing \cite{chuang} and much effort has been done, both
theoretically and experimentally, in order to achieve a deeper understanding of
this property. For a non-zero interaction between the system under
consideration and the surrounding environment, information about the system
will decohere into the environment, which is one of the main limitations for
scalable quantum computing. However, the last two decades have seen great
progress in achieving isolated controllable quantum mechanical systems.
Cavity quantum electrodynamics provides one such system; an atom is coupled
to a quantized electromagnetic field through a dipole-interaction inside a
high-Q cavity. The simplest example is described by the Jaynes-Cummings
model \cite{jc}, where a two-level atom interacts with just one mode of the
cavity field. This model is, within the rotating wave approximation,
analytically solvable and has been widely investigated since it was first
introduced. In spite of its simplicity, many interesting purely quantum
mechanical phenomena, such as, for example, revivals of Rabi oscillations and
field squeezing, may be understood from it. It is, in most cases, easy to
generalize the Jaynes-Cummings model to other similar systems, for example,
multi-mode fields and multi-level atoms, see the review article
\cite{genjc}. Some of these generalized models can still be solved
analytically and in this paper we focus on one of them.

We consider a three-level $\Lambda$-type atom coupled to two non-degenerate
cavity modes. By tuning the two transitions off resonance, the upper atomic
level can be adiabatically eliminated and the atomic system may be described
by an effective two-level atom. The lower atomic levels are thus coupled to
each other through the two modes, and in general the three subsystems,
mode-mode-atom, become highly entangled. This entanglement can persist even
when the photon numbers become large and we would otherwise enter a
classical regime.

For the various generalized Jaynes-Cummings models, the classical limit is
when the initial cavity modes of interest are prepared in coherent states
with a large average number of photons. The behavior of the ordinary
Jaynes-Cummings model in this limit was investigated in great detail in
\cite{GB} and it was found that, at particular interaction times, the atom
disentangled from the field, independently of its initial atomic state. At
these disentanglement times the field is in a so-called Schr\"odinger cat
state, that is a superposition of two states macroscopically far apart. The
interaction splits the initial coherent state into two parts in phase space,
which at the disentanglement times differ in phase by an angle $\pi$. The
theory outlined in \cite{GB} has recently also been demonstrated
experimentally \cite{haroche}. These Schr\"odinger cat states are well
suited for the study of how decoherence depends on the "size" of the system,
quantum information is supposed to be lost more rapidly to the environment
the more classical the system is. The quantum mechanical superposition then
collapses into a statistical mixture of the two parts,
see reference \cite{collapse}.

The classical limit in our model has been investigated in \cite{kines}.
However, reference \cite{kines} did not investigate the generation of
particular states of the two modes, but looked at the amount of
mode-entanglement in special cases. We will show that, for some initial
states of the two modes with large average of photons, the atom will
disentangle from the modes and that the modes are in entangled Schr\"odinger
cat states at these times. These states are in one sense more non-classical
than the Schr\"odinger cat states generated in the ordinary Jaynes-Cummings
model, since the two modes are also entangled with each other. These states
may be used, for example, to investigate decoherence and also to check
`non-locality' of quantum mechanics. We find that the disentanglement times
and the form of the Schr\"odinger cat states depend on the ratio between the
average number of photons in the two modes. A consequence of this is that
the disentanglement time may be small in the classical limit, which differs
from the ordinary Jaynes-Cummings model where this time goes as the square
root of the average of photons. So, the theory predicts that the atom will
disentangle from the modes at a certain time and that the modes and the atom
are then in some specific states. Thus, by measuring the atomic state at
this time we achieve a direct confirmation that the process of generating
entangled Schr\"odinger cat states was successful. The outcome of the atomic
measurement gives us a check, and makes the preparation scheme more robust.

In section \ref{sec:model} we present the model and give the general
analytical solution and also some of its properties. In the following
section \ref{disent}, the ideas from reference \cite{GB} about the classical
limit are applied to our system: we then derive the atomic disentanglement
times and approximate expressions for the field state. The validity of the
approximations are also investigated by calculating the purity of the
separate modes. It is found that initial coherent states are not sharp
enough, in terms of uncertainties in photon numbers, in order to prepare the
entangled Schr\"odinger cat states. In section \ref{qi} we give examples of
several quantum logic gates and examples of entangled states achievable
through the model. Finally we conclude the paper in section \ref{concl} with
a summary.

\section{The model}
\label{sec:model}

We consider a three-level $\Lambda$-atom, with lower levels
$|a\rangle$, $|b\rangle$ and upper level $|c\rangle$, interacting
with two non-degenerate microwave modes. The energies of the
respective atomic levels and field modes are:
$\hbar\Omega_{a,b,c}$ and $\hbar\omega_{1,2}$. The states
$|b\rangle$ and $|a\rangle$ couple to the state $|c\rangle$
through a dipole interaction with modes one and two respectively.
In the situation
\begin{equation}
\hbar\Delta=\hbar(\Omega_c-\Omega_b)-\hbar\omega_1=\hbar(\Omega_c-\Omega_a)-\hbar\omega_2,
\end{equation}
see figure 1, the population in state $|c\rangle$ can be
adiabatically eliminated provided that the detuning $\Delta$ is
large; see reference \cite{bose}. The atom-field evolution, after the
elimination, is then governed by the effective Hamiltonian
($\hbar=1$)
\begin{equation}\label{effham}
H=g(a_2^{\dagger}a_1\sigma^-+a_2a_1^{\dagger}\sigma^+),
\end{equation}
in the dipole and the rotating wave approximations. Here the
$a_{1,2}$'s are the boson operators for the two modes,
$\sigma^+=|b\rangle\langle a|$, $\sigma^-=|a\rangle\langle b|$ and
the effective coupling is $g=g_{ac}g_{bc}/\Delta$ with $g_{ij}$
being the original dipole couplings between the corresponding
atomic levels (here $g_{ab}=0$). With the Hamiltonian
(\ref{effham}), the number of excitations in the two modes,
$N=a_1^{\dagger}a_1+a_2^{\dagger}a_2$, is clearly conserved. This
symmetry can be used, for example, to calculate the atomic inversion
for special cases, as is shown below.

\begin{figure}[ht]
\centerline{\includegraphics[width=8cm]{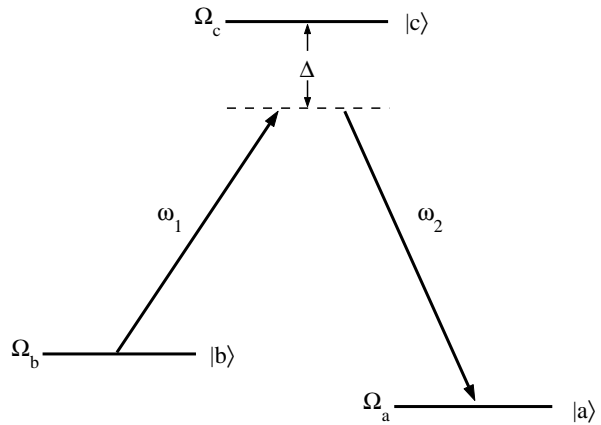}}
\caption{Energy level diagram of the three-level $\Lambda$-atom. The detuning
  $\Delta$ is assumed to be large so that the atomic level $|c\rangle$ can
  be eliminated. The two mode frequencies are represented by $\omega_1$ and
  $\omega_2$.  }
\label{fig1}
\end{figure}

A general disentangled initial state of the system, excluding the upper
state $|c\rangle$, can be written as
\begin{equation}
|\Psi(0)\rangle=\sum_{n,m}C_n^{(1)}C_m^{(2)}|n,m\rangle\left[\gamma|a\rangle+\delta|b\rangle\right],
\end{equation}
where $|n,m\rangle$ refers to $n$ photons in mode  one and $m$ photons in mode two. After a
time $t$ the state will evolve into
\begin{equation}\label{stateevo}
\begin{array}{ccl}
|\Psi(t)\rangle & = &
\sum_{n,m}C_n^{(1)}C_m^{(2)}\left\{\gamma\left[\cos(gt\sqrt{(n+1)m})|n,m,a\rangle\right.\right.
\\ \\ & &  \left.-i\sin(gt\sqrt{(n+1)m})|n+1,m-1,b\rangle\right]
\\ \\ & &
+\delta\left[\cos(gt\sqrt{(m+1)n})|n,m,b\rangle\right. \\ \\ & &
\left.\left.-i\sin(gt\sqrt{(m+1)n})|n-1,m+1,a\rangle\right]\right\}.
\end{array}
\end{equation}
Here it is understood that the states
$|n,m,a\rangle\equiv|n,m\rangle|a\rangle$ 
and $|n,m,b\rangle\equiv|n,m\rangle|b\rangle$.

Using the above expression (\ref{stateevo}), the atomic inversion
becomes
\begin{equation}\label{inversion}
\begin{array}{lll}
W(t) & = & \langle\Psi(t)|\sigma_z|\Psi(t)\rangle \\ \\ & = &
\sum_{n,m}\left[|C_n^{(1)}|^2|C_m^{(2)}|^2|\gamma|^2\cos^2(gt\sqrt{(n+1)m})\right.
\\ \\ & & +|C_{n+1}^{(1)}|^2|C_{m-1}^{(2)}|^2|\delta|^2\sin^2(gt\sqrt{(n+1)m})
\\ \\ & & -|C_n^{(1)}|^2|C_m^{(2)}|^2|\delta|^2\cos^2(gt\sqrt{(m+1)n})
\\ \\ & & \left.-|C_{n-1}^{(1)}|^2|C_{m+1}^{(2)}|^2|\gamma|^2\sin^2(gt\sqrt{(m+1)n})\right].
\end{array}
\end{equation}
An interesting observation is that if the photon-distributions of the two
modes is identical, $|C_n^{(1)}|=|C_m^{(2)}|,\,\,\,n=m$, the inversion is
strictly zero in the case when $|\gamma|=|\delta|$.  This does not, however,
mean that the atom is not entangled with the field-modes, as we will see in
the next section. This phenomenon, as mentioned above, is a consequence of
the symmetry in the system; for equal field intensities, population transfer
is equally likely in both directions between the two atomic states.  The
ordinary Jaynes-Cummings model does not have this symmetry.  Note that we
have assumed the two modes to be initially disentangled, but the result
holds also for entangled modes as long as the two modes have equal
photon-distributions. Another example of atomic population trapping in the
Raman model is studied in \cite{atomtrap}. There, one particular state of
the field is found, such that the inversion $W(t)$ is constant for any
initial atomic state.

\section{Evolution for large initial fields}\label{disent}
\subsection{Atom-field disentanglement}\label{afdis}

Following the method in \cite{GB} we now investigate the behavior of the
combined atom-field system when both the initial fields of the modes are
large. This situation has been investigated in reference \cite{kines}.
However, in reference \cite{kines} only some specific cases of the evolution of
the system where studied, and reference \cite{kines} did not discuss
how it may be used for state preparation.

We assume the two photon number distributions to be peaked
around the average photon numbers $\bar{n}$ and $\bar{m}$ and
further, in the large photon limit, $\bar{n},\bar{m}\gg 1$, we
replace $C_{n\pm1}^{(1)}C_{m\mp1}^{(2)}$ by $C_n^{(1)}C_m^{(2)}e^{i\varphi_{nm}^{(\pm)}}$.
We also put the phase to be constant,
$\varphi_{nm}^{(\pm)}=\pm\varphi$, under the assumption that the
phases of the two fields are slowly varying around the average
photon numbers. For large initial fields, the state
(\ref{stateevo}) can then, with the above approximations, be
written as
\begin{equation}\label{approx1}
\begin{array}{lll}
|\Psi(t)\rangle & \approx &
\sum_{n,m}C_n^{(1)}C_m^{(2)}\left\{\left[\gamma\cos(gt\sqrt{(n+1)m})\right.\right. \\ \\  & &
\left.-i\delta e^{i\varphi}\sin(gt\sqrt{(n+1)m})\right]|n,m,a\rangle \\ \\
& + & \left[\delta\cos(gt\sqrt{(m+1)n})\right.\\ \\ & & \left.\left.-i\gamma
e^{-i\varphi}\sin(gt\sqrt{(m+1)n})\right]|n,m,b\rangle\right\}.
\end{array}
\end{equation}
We introduce the new orthogonal atomic basis states
$|\phi_{\pm}\rangle=\frac{1}{\sqrt{2}}(e^{i\varphi}|a\rangle\pm|b\rangle)$,
and using (\ref{approx1}) with $\gamma=\frac{1}{\sqrt{2}}e^{i\varphi}$ and Ê$\delta=\frac{1}{\sqrt{2}}$ we find for these states
\begin{equation}\label{approx2}
\begin{array}{lll}
|\Psi(t)\rangle_{\pm} & \approx &
\frac{1}{\sqrt{2}}\sum_{n,m}C_n^{(1)}C_m^{(2)}e^{\mp
igt\sqrt{(m+1)n}}\times \\ \\ & & \left[e^{i\varphi}e^{\pm
igt\left[\sqrt{(m+1)n}-\sqrt{(n+1)m}\right]}|a\rangle\pm|b\rangle\right]|n,m\rangle
\\ \\ & \approx &
\frac{1}{\sqrt{2}}\left(e^{i\varphi}e^{\pm
i\frac{gt}{2}\left(\sqrt{\frac{\bar{n}}{\bar{m}}}-\sqrt{\frac{\bar{m}}{\bar{n}}}\right)}|a\rangle\pm|b\rangle\right)\times \\ \\ & & \sum_{n,m}C_n^{(1)}C_m^{(2)}e^{\mp
igt\sqrt{(m+1)n}}|n,m\rangle \\ \\ & = &
\frac{1}{\sqrt{2}}\left(e^{i\varphi}e^{\pm
i\frac{gt}{2}\left(\frac{\kappa-1}{\sqrt{\kappa}}\right)}|a\rangle\pm|b\rangle\right)|\psi_{\mp}\rangle.
\end{array}
\end{equation}
In the second step we have made the approximation
\begin{equation}
\sqrt{(m+1)n}-\sqrt{(n+1)m}\approx\frac{1}{2}\left(\sqrt{\frac{\bar{n}}{\bar{m}}}-\sqrt{\frac{\bar{m}}{\bar{n}}}\right),
\end{equation}
assuming the photon-distributions to be highly peaked around their
averages and that $\bar{n},\bar{m}\gg 1$. The last step defines
the field states $|\psi_{\pm}\rangle$ and the parameter
$\kappa=\bar{n}/\bar{m}$ as the ratio between the average photon
numbers of the two modes.

For effective interaction times
\begin{equation}\label{distime}
\begin{array}{lc}
gt_0^{(j)}=(2j+1)\frac{\pi\sqrt{\kappa}}{|\kappa-1|}, & j=0,1,2,...
\end{array}
\end{equation}
the two atomic parts in equation (\ref{approx2}) become equal. Thus,
since any initial atomic state can be written as a linear
combination of the two orthogonal states $|\phi_{\pm}\rangle$, the
atomic state will always disentangle from the field state at these
times. So, the interaction, in the large photon limit, describes a
non-unitary evolution in the Hilbert-space of the atom
\cite{GB}. However, this disentanglement between the atom and the
field,
$|\Psi(gt=gt_0^{(j)})\rangle=|\psi_{atom}\rangle\otimes|\psi_{field}\rangle$,
only occurs if $\kappa\neq 1$. Note, by adjusting the ratio
between the average photon numbers in the two modes, varying
$\kappa$, it is possible to control the disentanglement time.

A measure of the degree of entanglement between different subsystems is the
purity defined as $P_A(t)=\mathrm{Tr}\left(\rho_A(t)^2\right)$. Here
$\rho_A(t)$ is the reduced density operator for system $A$, obtained by
tracing out the other subsystems' degrees of freedom from the full system's
density operator, $\rho_A(t)=\mathrm{Tr}_B\left(\rho(t)\right)$. In
reference \cite{GB}, for the ordinary Jaynes-Cummings model, the field was
assumed to be in coherent states. However, for a sharper distribution the
agreement between the large field approximations and the exact results is
supposed to be enhanced, which is also shown in \cite{knight}. We will
assume the initial states of the two modes to be either in coherent or
squeezed coherent states. For a coherent state
$|\nu\rangle=D(\nu)|0\rangle$, where $D(\nu)=\exp\left(\nu
a^{\dagger}-\nu^*a\right)$ is the displacement operator, we have
\begin{equation}
C_n^{(1)}=e^{-|\nu|^2/2}\frac{\nu^n}{\sqrt{n!}}
\end{equation}
and similarly for mode two
\begin{equation}
C_m^{(2)}=e^{-|\mu|^2/2}\frac{\mu^m}{\sqrt{m!}}.
\end{equation}
For a squeezed coherent state $|r,\nu\rangle=S(r)|\nu\rangle$,
where $S(r)=\exp\left(r(a^2-a^{\dagger2})/2\right)$, the
coefficients are given by \cite{knight}
\begin{equation}\label{sqs}
C_n^{(1)}\!=\!\frac{\tanh(r)^{n/2}}{\sqrt{n!\,2^n\cosh(r)}}e^{-|\nu|^2(1-\tanh(r))/2}H_n\!\!\left(\frac{\nu}{\sqrt{\sinh(2r)}}\right)\!,
\end{equation}
if $\nu$ is chosen real and $H_n$ is the $n$'th Hermite polynomial and, of
course, we have a similar result for mode two. In the following, for
convenience, we always choose the phases of the initial fields to be zero.
Figure 2 shows the atomic purity, $\mathrm{Tr}\left(\rho_A(t)^2\right)$, as a function of the effective interaction
time, both for initial coherent states (dashed curve) and squeezed coherent
states (solid line). The reduced atomic density operator is achieved from equation (\ref{stateevo}), by constructing the full systems density operator and then trace over the two modes degrees of freedom, as mentioned above. In both cases we have $\bar{n}=|\nu|^2=150$ and
$\bar{m}=|\mu|^2=50$, the squeezing parameters for the solid curve are
$r_1=r_2=1$ for both modes and the initial atomic state is $|a\rangle$. It
is clear that in the squeezed case, when the photon-distribution is sharper
(sub-Poissonian), the disentanglement of the atom is improved. This is most
prominent for larger disentanglement times $gt_0^{(j)}$. For $j=0$ the
atomic purity is almost the same for the two cases; solid and dashed curves.
The additional peaks at around $gt\approx11$ and $gt\approx22$, seen in the figure, are
due to the disentanglement at the revival times; see reference
\cite{zaheer}. This phenomenom is not predicted by the theory given above,
see equation (\ref{distime}).

\begin{figure}[ht]
\centerline{\includegraphics[width=8cm]{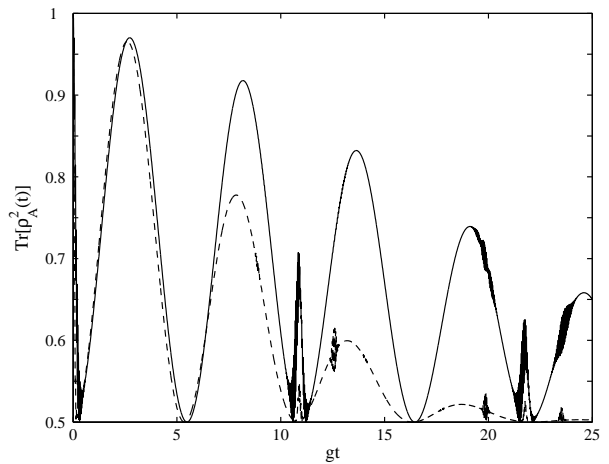}}
\caption{
  This figure shows the atomic purity as a function of the effective
  interaction time when the atom is initially in the state $|a\rangle$ and two modes either in coherent states (dashed curve) or squeezed states (solid curve). The average numbers of
  photons are $\bar{n}=150$ and $\bar{m}=50$ in both cases and the squeezing
  parameters are $r_1=r_2=1$, see equation (\ref{sqs}). The atom disentangles
  from the field when the purity equals unity. It is clear the the amount of
  disentanglement improves for the sub-Poissonian squeezed states,
  especially at larger times. The sharp peaks at around $gt\approx 11$ and
  $gt\approx 22$ are the separation of the atom from the field at the
  revival times.}
\label{fig2}
\end{figure}

\subsection{Field states at the disentanglement time}

We have seen that at certain times $gt_0^{(j)}$ given by equation
(\ref{distime}), the state of the whole system can be written as a
product of the atomic state and the field state in the large
photon approximation. The state of the two modes will at these
times be a linear superposition of the states
$|\psi_{\pm}\rangle$, where the coefficients are determined from
the initial atomic state. For an atom initially in the state
$|a\rangle$ we obtain, for example,
\begin{equation}
|\psi_{field}(gt_0^{(j)})\rangle=\frac{1}{N}\left(|\psi_+(gt_0^{(j)})\rangle+|\psi_-(gt_0^{(j)})\rangle\right),
\end{equation}
where $N$ is a normalization constant.

As in section \ref{afdis} we try to make use of the sharpness of the
photon-distributions when the average photon numbers becomes large
in order to approximate the expressions for the field states
$|\psi_{\pm}\rangle$. If we assume the fields to be initially in
coherent states and expand $\sqrt{(m+1)n}$ to first order around
$\bar{n}$ and $\bar{m}$ and make use of $\bar{n},\bar{m}\gg 1$,
the states become
\begin{equation}\label{fieldapprox}
|\psi_{\pm}(t)\rangle\approx e^{\pm
i\frac{\sqrt{\kappa}gt}{2}}|\nu e^{\pm
i\frac{gt}{2\sqrt{\kappa}}}\rangle|\mu e^{\pm
i\frac{\sqrt{\kappa}gt}{2}}\rangle.
\end{equation}
At the first disentanglement time $gt_0^{(0)}$ we get
\begin{equation}\label{fieldapprox2}
|\psi_{\pm}(gt_0^{(0)}\rangle\approx e^{\pm
i\frac{\pi\kappa}{2|\kappa-1|}}|\nu e^{\pm
i\frac{\pi}{2}\frac{1}{|\kappa-1|}}\rangle|\mu e^{\pm
i\frac{\pi}{2}\frac{\kappa}{|\kappa-1|}}\rangle.
\end{equation}
It is interesting to note that the difference in phase between the
two modes is independent of $\kappa$ at the disentanglement times:
$(2j+1)\frac{\pi}{2}\frac{\kappa}{|\kappa-1|}-(2j+1)\frac{\pi}{2}\frac{1}{|\kappa-1|}=\pm(2j+1)\frac{\pi}{2}$, where the $\pm$ sign depends on wether $\kappa$ is larger or smaller than 1  
(the phases of the initial fields are set equal to zero). If
higher order terms were included in the approximation, the
exponentials would have contained mixing terms of the form
$n^pm^q$, where $p,q=1,2,3,...\,$. These terms have the effect of
entangling the two modes, and thus, for large times $gt$, when the
higher order terms can not be neglected, the above approximation
must fail.

From the approximate states (\ref{fieldapprox}) it is easy to get
an expression for the revival times by setting the two states
equal and solving for $t$
\begin{equation}\label{revival}
|\nu e^{i\frac{gt_r}{2\sqrt{\kappa}}}\rangle|\mu
e^{i\frac{\sqrt{\kappa}gt_r}{2}}\rangle=|\nu
e^{-i\frac{gt_r}{2\sqrt{\kappa}}}\rangle|\mu
e^{-i\frac{\sqrt{\kappa}gt_r}{2}}\rangle 
\end{equation}
leading to
\begin{equation}
\begin{array}{cc}\left\{
\begin{array}{l} gt_r=2\pi\sqrt{kl} \\
\kappa=l/k
\end{array}\right., & l,k=1,2,3,...
\end{array}.
\end{equation}

The above expression holds for large initial coherent field
states, but a similar behavior is expected for other sharp highly
excited field states, such as, for example, sub-Poissonian states. For
small $\bar{n}$ and $\bar{m}$ the revival times (\ref{revival})
seem still to hold, but secondary revivals between the above times
occur, see \cite{revival}.

Due to the exponent $\exp(\pm igt\sqrt{(m+1)n})$ in the expression
for the states $|\psi_{\pm}(t)\rangle$ we see that the two modes
in these states, as mentioned above, will become entangled for
large times $gt$. The degree of entanglement will monotonously
increase with $gt$ until the two modes becomes maximally
entangled. This is shown in figure 3 where the purity for the modes,
$P_{mode}^{\pm}(t)=\mathrm{Tr}_{1}\left[(\rho_{1}^{\pm})^2\right]
=\mathrm{Tr}_{2}\left[(\rho_{2}^{\pm})^2\right]$,
is plotted. (We must, of course, have that
$P_{mode}^+=P_{mode}^-$.) The reduced density operators for the
modes are
$\rho_{1,2}^{\pm}=\mathrm{Tr}_{2,1}\left(|\psi_{\pm}\rangle\langle\psi_{\pm}|\right)$,
where the indices refer to the two modes. The states are the same
as in figure 2, $\bar{n}=150$, $\bar{m}=50$ and the squeezing
parameters $r_1=r_2=1$. The dashed curve corresponds to initial
coherent states while the solid curve is with squeezed states. The
sub-Poissonian squeezed states stay disentangled for a longer
time than the Poissonian coherent states as expected.

\begin{figure}[ht]
\centerline{\includegraphics[width=8cm]{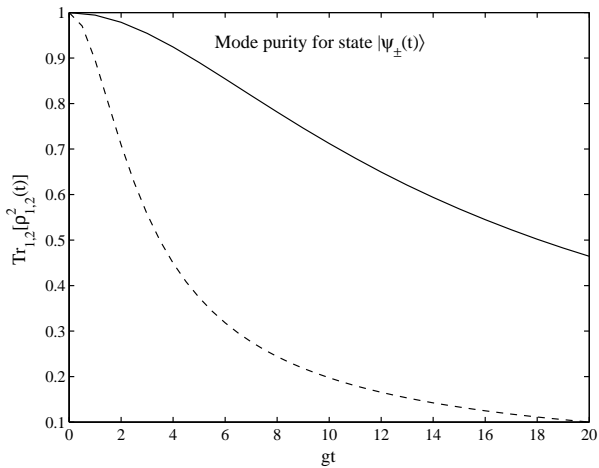}}
\caption{
  This shows the mode purity as function of the effective interaction time
  for the field states $|\psi_{\pm}(t)\rangle$, defined in equation
  (\ref{approx2}). The solid curve is the purity for initial squeezed states
  and the dashed curve when the two modes are prepared in coherent states.
  For both cases, the average numbers of photons are $\bar{n}=150$ and
  $\bar{m}=50$ and the squeezing parameters are $r_1=r_2=1$. For large times
  the two modes becomes entangled, which is because of the exponent
  $\exp(\pm igt\sqrt{(m+1)n})$ in the expression for
  $|\psi_{\pm}(t)\rangle$. The mixing of the modes is greatly reduced for
  the squeezed states due to the sharp photon distribution.}
\label{fig3}
\end{figure}

Even while the two modes in the states $|\psi_{\pm}\rangle$
becomes entangled, it is not necessarily the case for the modes in
the actual state of the field. We saw that at the disentanglement
times, the field will be in a linear combination of
$|\psi_+\rangle$ and $|\psi_-\rangle$, also if we do a conditional
measurement of the atomic state, after a time $t$, the field will
be left in a linear combination of the above states. Assume, for
example, an atom, initially in the state $|b\rangle$, also being
measured in $|b\rangle$ after a time $t$, the field will then be
in the state
\begin{equation}
\begin{array}{lll}
|\psi_{field}(t)\rangle & = & \frac{1}{N}\sum_{n,m}C_nC_m\cos(gt\sqrt{(m+1)n})|n,m\rangle \\ \\ 
& = & \frac{1}{2N}\left(|\psi_+(t)\rangle+|\psi_-(t)\rangle\right),
\end{array}
\end{equation}
where $N$ is a normalization constant. The mode purity for this
state is plotted in figure 4 with $\bar{n}=100$, $\bar{m}=50$ and
$r_1=r_2=1$. At the disentanglement times $gt_0^{(j)}$ the purity
is greatly increased. At these times, according to equation
(\ref{fieldapprox}), the state of mode 2 becomes equal in both of
the states $|\psi_{\pm}\rangle$ and consequently disentangled from
mode 1.

\begin{figure}[ht]
\centerline{\includegraphics[width=8cm]{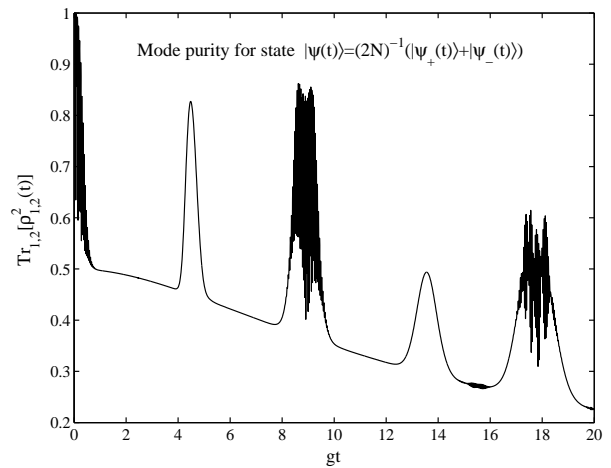}}
\caption{
  This, as in the previous figure 3, also gives the mode purity, but for the
  field state
  $|\psi(t)\rangle=(2N)^{-1}(|\psi_+(t)\rangle+|\psi_-(t)\rangle)$, where
  $|\psi_{\pm}(t)\rangle$ are defined in equation (\ref{approx2}). Here $N$
  is a normalization constant and the two modes are initially in squeezed
  states (\ref{sqs}) with $\bar{n}=100$ and $\bar{m}=50$ and squeezing
  parameters $r_1=r_2=1$. The periodic increase of the purity can be
  understood from equation (\ref{fieldapprox}). At these times, either the
  state of mode one or two, becomes equal in the approximate expressions for
  the two states $|\psi_{\pm}(t)\rangle$ and consequently disentangle from
  the other mode.}
\label{fig4}
\end{figure}

\section{Quantum information processing and entangled states}\label{qi}
In this section we show how the Raman-coupled model could be used
for quantum logic operations and how various entangled states of
the field and atoms can be prepared. In all these examples, except in
section \ref{cat}, the fields just contain a few photons and thus,
the large field approximations are not used. However, in the
section \ref{cat} on preparation of Schr\"odinger cat states of the
two modes, the large field approximations are very helpful for
getting a better understanding of the dynamics.

\subsection{Field mode quantum logic}
Let us assume that the effective interaction time is chosen
$gt\sqrt{3}=\pi$ and that the two modes are prepared in either of
the Fock states $|0\rangle$ or $|3\rangle$. For the atom initially
in $|a\rangle$ or $|b\rangle$, the state will, according to equation (\ref{stateevo}), evolve as
\begin{equation}\label{phgate}
\begin{array}{lcl}
|0,0,a\rangle\rightarrow|0,0,a\rangle & &
|0,0,b\rangle\rightarrow|0,0,b\rangle \\ \\
|0,3,a\rangle\rightarrow-|0,3,a\rangle & &
|0,3,b\rangle\rightarrow|0,3,b\rangle \\ \\
|3,0,a\rangle\rightarrow|3,0,a\rangle & &
|3,0,b\rangle\rightarrow-|3,0,b\rangle \\ \\
|3,3,a\rangle\rightarrow|3,3,a\rangle & &
|3,3,b\rangle\rightarrow|3,3,b\rangle.
\end{array}
\end{equation}
The above transformation describes a quantum phase gate on the two
modes, where the atom state acts as an ancilla state. Note that the
initial state of the atom decides which field state that changes
sign.

\subsection{Atomic quantum logic}
Having looked at quantum logic with field modes acting as qubits, we now
show how the atom also can be used as an information carrier. Assume mode
one to be in either of the Fock states $|0\rangle$ or $|n'\rangle$ while
mode two is always in vacuum, $|0\rangle$. If the interaction time is such
that $gt\sqrt{n'}=\pi$, we get for the atomic states
$|\phi_{\pm}\rangle=\frac{1}{\sqrt{2}}(|a\rangle\pm|b\rangle)$
\begin{equation}\label{atomcnot}
\begin{array}{c}
|0,0,\phi_{\pm}\rangle\rightarrow|0,0,\phi_{\pm}\rangle \\ \\
|n',0,\phi_{\pm}\rangle\rightarrow|n',0,\phi_{\mp}\rangle.
\end{array}
\end{equation}
Thus, the atomic state is flipped if mode one is in the state
$|n'\rangle$ while it is unchanged if mode one is in the vacuum.
The transformation (\ref{atomcnot}) is identified as a C-NOT quantum
logic gate; mode one is the control bit and the atom, the target
bit, is flipped depending on the state of the control bit.

\subsection{Entangled Schr\"odinger cat states}\label{cat}

In section \ref{disent} it was shown that when the atom
disentangles from the field, in the large photon limit, the state
for the modes will be linear combination of the states
$|\psi_{\pm}(gt_0^{(j)})\rangle$. If the initial state of the atom
is $|a\rangle$ or $|b\rangle$ the weight of these field states
will be the same. In a first approximation, for coherent field
states, the interaction modifies the states by just a phase. If the
initial field phases are zero, it means that the interaction will
split up the initial state into two parts, one with negative and
one with positive phases. For example, an initial state
$|\Psi(0)\rangle=|\nu,\mu,a\rangle$ 
and choosing $\kappa=3$ ($\kappa=\bar{n}/\bar{m}$) will evolve, using the approximation
(\ref{fieldapprox2}), into
\begin{equation}\label{cat1}
\begin{array}{lll}
|\Psi(gt_0^{(0)})\rangle& = & \frac{1}{\sqrt{2}}\left(i|a\rangle+|b\rangle\right)\otimes \\ \\ & &\!\! \frac{i}{N}\!\!\left(\!|\nu
e^{-i\pi/4}\rangle|\mu e^{-i3\pi/4}\rangle-|\nu
e^{i\pi/4}\rangle|\mu e^{i3\pi/4}\rangle\!\right)\!,
\end{array}
\end{equation}
where, again, $N$ is the normalization constant. Thus, the two
field modes in the above expression are both in Schr\"odinger cat
states and they are, at the same time, also maximally entangled
with each other. However, for initial coherent states, the two
modes in the states $|\psi_{\pm}\rangle$ will become entangled
with each other very rapidly, so that the approximation
(\ref{fieldapprox2}) fails for disentanglement times not short
enough. From the figure 3 (dashed curve) we note that the purity of the
two modes at the first disentanglement time is $P_{mode}\approx 0.60$
for $\bar{n}=150$ and $\bar{m}=50$. So the expression (\ref{cat1})
for the state is in fact not a good approximation when $\bar{n}=150$
and $\bar{m}=50$. On the other hand, we saw that the time-scale for mode-mode
entanglement was much longer for sub-Poissonian squeezed coherent
states. This suggests that by using squeezed coherent initial
field states, it is possible to prepare the two modes in entangled
Schr\"odinger cat states. According to figure 3 (solid curve) the
corresponding mode purity is $P_{mode}\approx 0.96$, which indicates
that the first order approximation (\ref{fieldapprox2}) is
supposed to be acceptable. Figure 5 is a plot of the
$Q$-functions,
$Q_{1,2}^{\pm}(\alpha,t)=\langle\alpha|\rho_{1,2}^{\pm}(t)|\alpha\rangle$,
for the two modes at times $t=0$ and $t=t_0^{(1)}$. The initial field states
are the same as the ones used for the solid curve in figure 3. The figure 5,
together with the purity $P_{mode}\approx 0.96$, clearly shows that the two
modes will be in an entangled Schr\"odinger cat state with the correct
estimated phases. Note that this kind of phase-space plot is not enough to
give all the information about the state of the two modes, for example, it
does not tell us the degree of entanglement between the modes. For long
times the large field approximations for the field states will eventually
fail, however, they still give an indication of how the separate states
behave in phase-space.

\begin{figure}[ht]
\centerline{\includegraphics[width=8cm]{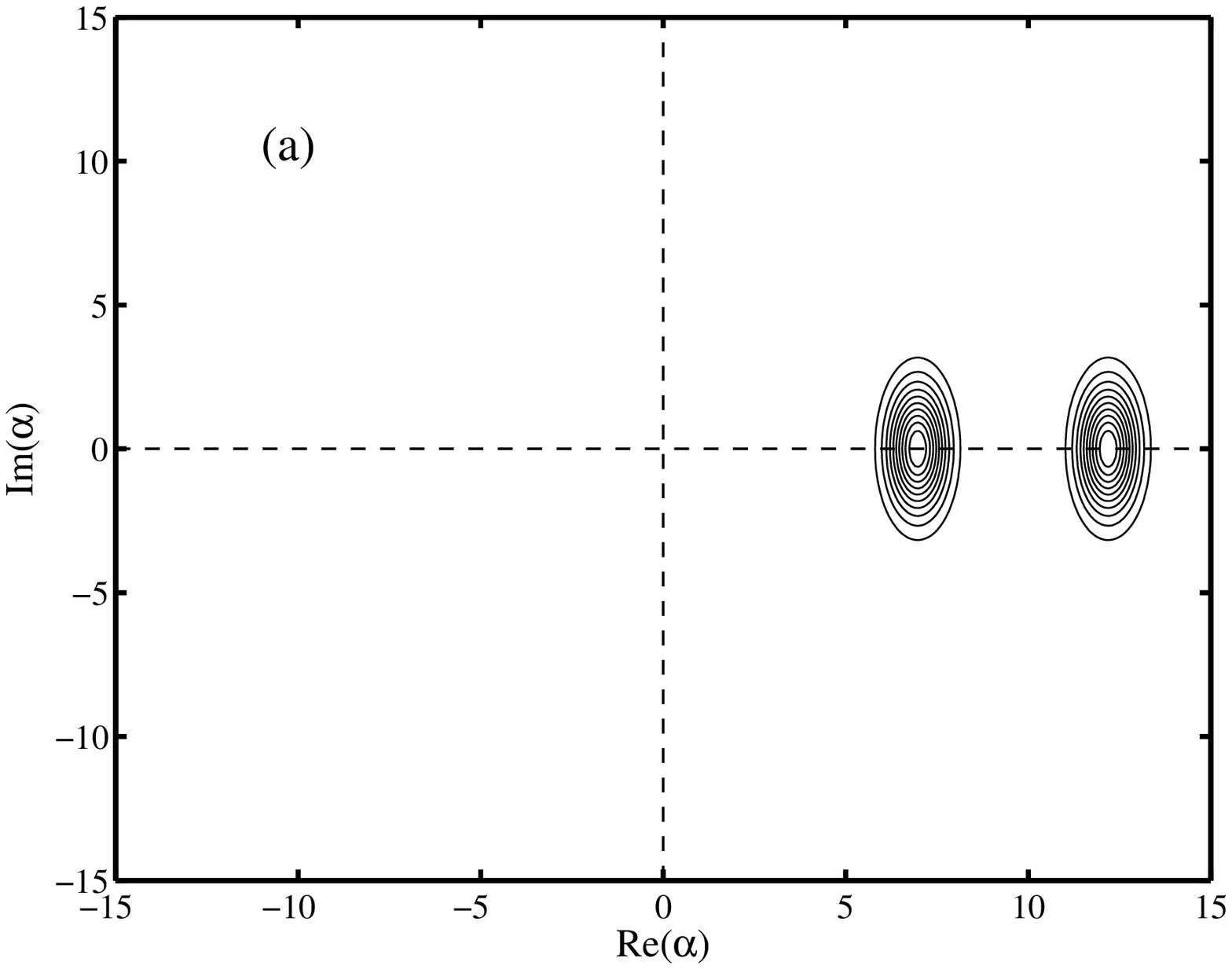}}
\centerline{\includegraphics[width=8cm]{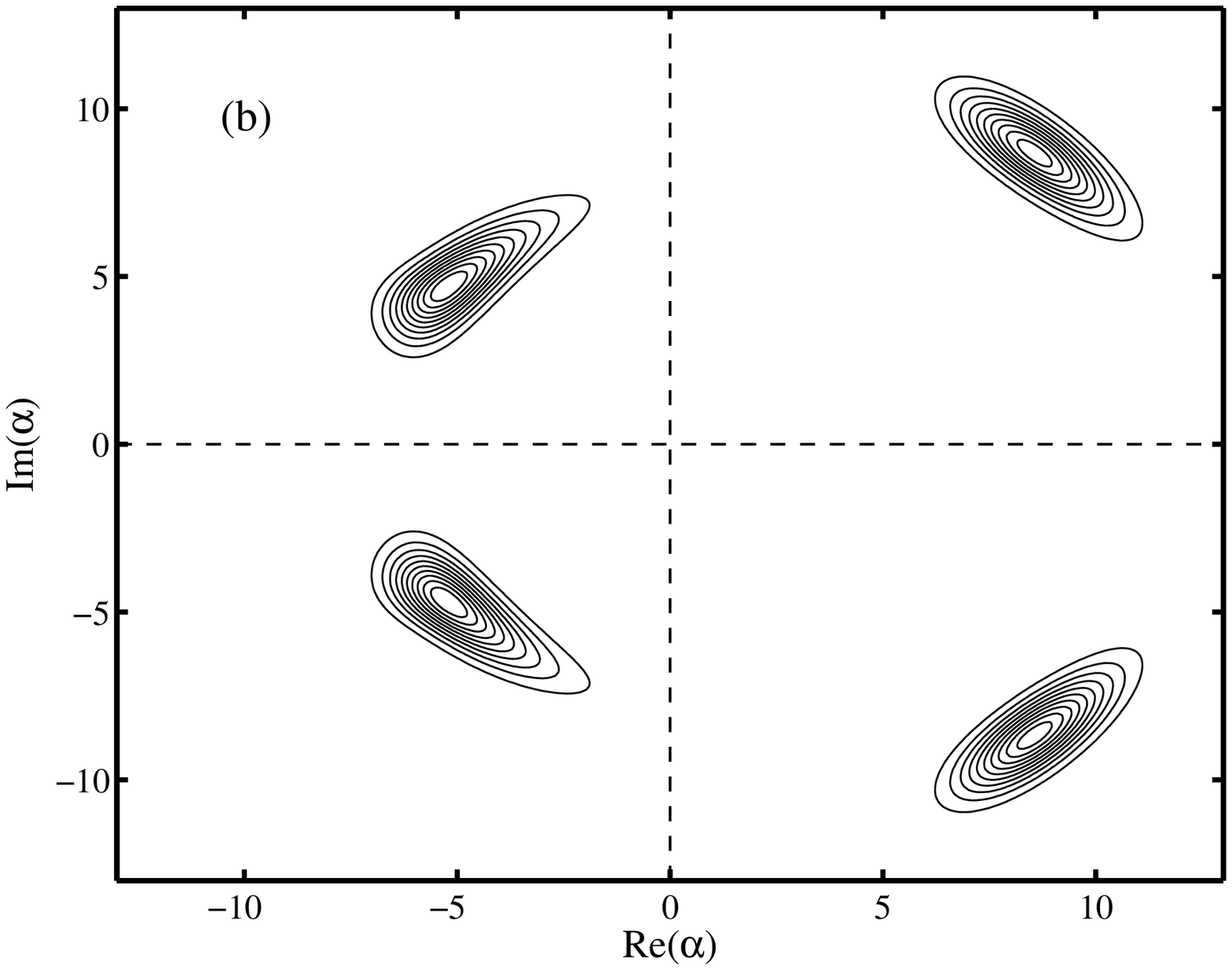}}
\caption{
  These two contour plots show the various reduced $Q$-functions
  $Q_{1,2}^{\pm}(\alpha,t)=\langle\alpha|\rho_{1,2}^{\pm}(t)|\alpha\rangle$,
  at (a) the time zero, and (b) first disentanglement time $gt_0^{(1)}$. The two
  modes are initially in squeezed states with $\bar{n}=150$, $\bar{m}=50$
  and $r_1=r_2=1$, which are seen in (a). In figure 5 (b) it is clear how
  the two separate modes' initial states splits up into two parts,
  characterizing the Schr\"odinger cats. Note that the phases of the fields
  agree well with the predicted phases from the approximations, see, for
  example, equation (\ref{cat1}). The shapes of the $Q$-functions for mode
  two, with lower average number of photons, have changed more during the
  evolution than for mode one.}
\label{fig5}
\end{figure}

\subsection{EPR-states}

For $gt=\pi/4$ we get, from equation (\ref{stateevo}), the following
evolution
\begin{equation}
\begin{array}{c}
|0,1,a\rangle\rightarrow\frac{1}{\sqrt{2}}\left(|0,1,a\rangle-i|1,0,b\rangle\right).
\end{array}
\end{equation}
If the atom, when it leaves the cavity after a time $gt=\pi/4$,
experiences a $\pi/2$-pulse, which means that
$|a\rangle\rightarrow\frac{1}{\sqrt{2}}(|a\rangle+|b\rangle)$ and
$|b\rangle\rightarrow\frac{1}{\sqrt{2}}(|a\rangle-|b\rangle)$, and
then the atomic state is measured in the
$\left\{|a\rangle,|b\rangle\right\}$ basis, the two will be in an
EPR-state. Depending on the measured atomic state, the two modes
are left in the maximally entangled states
\begin{equation}
\begin{array}{c}
|\mathrm{EPR}\rangle_-=\frac{1}{\sqrt{2}}(|0,1\rangle-i|1,0\rangle)
\\ \\
|\mathrm{EPR}\rangle_+=\frac{1}{\sqrt{2}}(|0,1\rangle+i|1,0\rangle),
\end{array}
\end{equation}
when the measurement result is $|a\rangle$ and $|b\rangle$
respectively.

\subsection{GHZ-states}

Let us assume that the two cavity modes could be prepared in the
states $|\pm\rangle=\frac{1}{\sqrt{2}}(|0\rangle\pm|3\rangle)$. We
again chose the interaction time so that $gt\sqrt{3}=\pi$ and we
see from (\ref{phgate}), that after two atoms has passed the
cavity, one after the other, the states of the modes will be
flipped if the two atomic states were $|a\rangle_1|b\rangle_2$ or
$|b\rangle_1|a\rangle_2$, while if both atoms where in the same
state during the passage, the field is left unchanged. For example
\begin{equation}
|+,+\rangle\!\frac{1}{\sqrt{2}}(|a\rangle_1\pm|b\rangle_1)|a\rangle_2\!\rightarrow\!\!\frac{1}{\sqrt{2}}(|+,+,a\rangle_1\pm|-,-,b\rangle_1)|a\rangle_2,
\end{equation}
which means that the interaction leaves the first atom and the two
cavity modes in a GHZ-state. Note that the order of the atoms does
not matter, so the prepared GHZ-state could equally well include
the two modes and the second atom instead.

\section{Conclusion}\label{concl}
In this paper we looked at a theoretical treatment of a three-level Raman
system which is coupled to a quantised two-mode cavity. Thus the system has
one atom and two modes as sub-systems. An adiabatic elimination of the upper
level ensures that when the atom transfers from one lower atomic state to
the other, there is a corresponding transfer of photons between the modes.
The atom passes through the cavity resulting in a finite interaction time,
but during that time we try to effect changes on the quantum states in the
cavity. By adjusting the interaction time, one of the three subsystems may
disentangle from the others, and can, in such a way, be used to perform a
controlled transformation of the remaining two systems. 

A key approximation in
our model, is concerned with the adiabatic elimination of the upper atomic
state, which requires a sufficient two-photon detuning. (For a calculation
of what is required, see Ref.~\cite{wu}.) Unfortunately,
such a detuning has the effect of reducing the effective two-photon coupling
$g$ when compared to the single photon couplings (such as $g_{ab}$).
This would slow down the dynamics and make the system more vulnerable to
decoherence. However, one advantage of generating entangled Schr\"odinger cat
states in this two-photon model is that it can be done very quickly, i.e.\ 
before decoherence becomes too significant. The new feature, compared to the
single mode Jaynes-Cummings model, is the appearance of the highly
controllable photon number ratio $\kappa$ appearing in the disentanglement
time, equation (\ref{distime}). 

We have shown how this Raman cavity
system can be used as a phase gate with the two field modes carrying qubits
and the atom acting as an ancilla state. The state of the atom can be
measured after this gate and be used to confirm the correct operation of the
phase gate without the destruction of the qubits. 

We have also shown how to
make a controlled-not gate if the atomic system holds a qubit. Recently many
schemes have been proposed for large scale quantum computation on trapped
atoms inside a cavity, see \cite{atomlogic}. The logic gate (\ref{atomcnot})
could be used in such models if external lasers, in a controlled way, are
used to Stark-shift the different energy levels for the atoms involved.We
have also shown how some interesting entangled states may be produced. In
particular, as well as entangled Schr\"odinger cat states, we can make
EPR-type states of the electromagnetic field and a GHZ state of the field
modes and an atom. These states and the proposed quantum logic elements may
make this an interesting system to pursue experimentally and theoretically
in the future.

\section*{Acknowledgements}

We would like to acknowledge support from the Royal Swedish
Academy of Sciences and the European Union (under
contract no.\ HMPT-CT-2000-00096).


\begin{thebibliography}{12}

\bibitem{chuang} Nielsen M. A., and Chuang I. L., 2000, \textit{Quantum Computing and Quantum Information} (Cambridge). 

\bibitem{jc} Jaynes, E. T., and Cummings, F. W., 1963, \textit{Proc. IEEE}, \textbf{51}, 89.

\bibitem{genjc} Messina A., Maniscalco S., and Napoli A., 2003, \textit{J. Mod. Opt.}, \textbf{50}, 1.

\bibitem{GB} Gea-Banacloche J., 1990, \textit{Phys. Rev. Lett.}, \textbf{65}, 3385, Gea-Banacloche J., 1991, \textit{Phys. Rev.} A, \textbf{44}, 5913.

\bibitem{haroche} Auffeves A., Maioli P., Meunier T., Gleyzes S., Nogues G., Brune M., Raimond J. M., and Haroche S., arXiv:quant-ph/0307185.

\bibitem{collapse} Walls D. F., and Milburn G. J., 1985, \textit{Phys. Rev.} A, \textbf{31}, 2403. Phoenix S. J. D., 1990, \textit{Phys. Rev.} A, \textbf{41}, 5132.

\bibitem{kines} Fang M. F., Zhou Q. P., and Zhou P., 1999, \textit{Acta Phys. Sin.}, \textbf{8}, 401.

\bibitem{bose} Alexanian M., and Bose S. K., 1995, \textit{Phys. Rev.} A,
\textbf{52}, 2218.

\bibitem{atomtrap} Deb B., Gangopadhyay G., and Ray S., 1993, \textit{Phys. Rev.} 
A, \textbf{48}, 1400.

\bibitem{knight} Rekdal P. K., Skagerstam B. S. K., and Knight P. L., arXiv:quant-ph/0301148.

\bibitem{zaheer} Nasreen T., and Zaheer K., 1994, \textit{Phys. Rev.} A, \textbf{49}, 616.

\bibitem{revival} Cardimona D. A., Kovanis V., Sharma M. P., and Gavrielides A., 1991, \textit{Phys. Rev.} A, \textbf{43}, 3710.

\bibitem{atomlogic} Feng M., 2002, \textit{Phys. Rev.} A, \textbf{66}, 054303, Pachos J., and Walther H., 2002, \textit{Phys. Rev. Lett.} \textbf{89},
187903-1.

\bibitem{wu} Wu Y.,  1996, \textit{Phys. Rev.} A, \textbf{54}, 1586.

\end{thebibliography}
\end{document}